\documentclass[hyper,12pt,a4paper]{article}
\usepackage{mathrsfs, subfigure}
\usepackage{amsmath}
\usepackage{amsfonts}
\usepackage{amssymb}
\usepackage{graphicx}

\usepackage{jheppub}

\newcommand{\be}{\begin{equation}}
\newcommand{\bea}{\begin{eqnarray}}
\newcommand{\eea}{\end{eqnarray}}
\newcommand{\ba}{\begin{array}}
\newcommand{\ea}{\end{array}}
\newcommand{\ee}{\end{equation}}

\def\bse{\begin{subequations}}
\def\ese{\end{subequations}}

\title{ Holographic SIS  Josephson Junction }

\author[a]{Yong-Qiang Wang,}
\author[a]{Yu-Xiao Liu,}
\author[b]{Rong-Gen Cai,}
\author[b]{Shingo Takeuchi,}
\author[b]{~~~Hai-Qing Zhang}

\affiliation[a]{Institute of Theoretical Physics, Lanzhou
University,  Lanzhou 730000,  China}
 \affiliation[b]{State Key
Laboratory of Theoretical Physics, Institute of Theoretical
Physics,\\ Chinese Academy of Sciences,  Beijing 100190, China.}

\emailAdd{yqwang@lzu.edu.cn} \emailAdd{liuyx@lzu.edu.cn}
\emailAdd{cairg@itp.ac.cn} \emailAdd{shingo@itp.ac.cn}
\emailAdd{hqzhang@itp.ac.cn}

\abstract{ We construct a holographic model for the
superconductor-insulator-superconductor (SIS) Josephson junction at
zero temperature by considering a complex scalar field coupled with
a Maxwell field in the four-dimensional anti-de
Sitter soliton background. From the gravity side we reproduce
the sine relation between the Josephson current and the phase
difference across the junction. We also study the dependence of the
maximal current on the dimension of the condensate operator and on
the width of the junction, and obtain expected results.}

\keywords{Holography, Josephson junction, AdS Soliton}

\preprint{CAS-KITPC/ITP-316}
\begin{document}
\maketitle

\section{Introduction}

 With the help of the AdS/CFT correspondence~\cite{Maldacena:1997re},
 the superconductivity and superfluid phenomena  have been studied
intensively on the gravity side in recent
years~\cite{Gubser:2008px,Hartnoll:2008vx,Hartnoll:2008kx}. For some
reviews, see, for example,
\cite{Hartnoll:2009sz,Herzog:2009xv,Horowitz:2010gk}.
 One of the interesting phenomena  associated with superconductivity
is the Josephson effect~\cite{Josephson:1962zz} (a brief description
of the Josephson effect is given in appendix
\ref{Josephsonjunction}), namely, the effect of electrons tunneling
between two superconductors separated by a weak link. The weak link
can be a normal conductor or an insulating barrier. Correspondingly
the Josephson junction is referred to as the
superconductor-normal-superconductor (SNS) or
superconductor-insulator-superconductor (SIS) junction,
respectively.

 Recently, a holographic model for a three-dimensional SNS
Josephson junction has been constructed and studied  by Horowitz
{\it et al} ~\cite{Horowitz:2011dz} in an AdS-Schwarzschild
black hole background with a Maxwell field and a complex
scalar field. The extension to  a four-dimensional Josephson
junction  has been discussed
in~\cite{Wang:2011rva,Siani:2011uj}. The Josephson junction array
based on a designer multigravity  has been  constructed in
\cite{Kiritsis:2011zq}, and the holographic p-wave Josephson
junction has also been  discussed in~\cite{Wang:2011ri}.
 In those works, some familiar features of Josephson
junction have been reproduced on the gravity side. Note that in
those studies the dual weak link is a normal conductor. Therefore it
would be of great interest to construct a holographic model
for a Josephson junction with an insulator link and study its
 features. This is the aim of the present paper.

In order to build the holographic model for an SIS
Josephson junction, we first need a model to realize
the superconductor/insulator phase transition. Fortunately
such a holographic model was constructed in~\cite{Nishioka:2009zj}.
There it was shown that in an Einstein-Maxwell-complex scalar field
theory with a negative cosmological constant, one can have two
phases.  One is the AdS soliton solution with a 
vanishing scalar field. This solution is dual to  a confined
gauge theory with a mass gap in the AdS boundary, which
resembles the insulator phase.  The other  is the AdS soliton
solution with non-vanishing scalar field.
 When chemical potential increases
beyond a certain value $\mu_c$, the above AdS soliton solution turns
out to be unstable, a new and stable AdS soliton solution with
nontrivial scalar field  appears. The new solution is dual to a
superconducting phase.  In this way one
realizes the superconductor/insulator phase transition.
The analytical study of the holographic insulator/superconductor
phase transition in a five-dimensional AdS soliton spacetime
was presented \cite{Cai:2011ky}.  With the holographic model
of superconductor/insulator phase transition,  we here construct a
holographic model dual to a  (1+1)-dimensional  SIS Josephson
junction in a four-dimensional Einstein-Maxwell-complex scalar
theory with a negative cosmological constant.

 In the probe limit, by numerically solving the coupled non-linear
equations of motion for the Maxwell field and scalar field in an AdS
soliton background, we show that the Josephson current is
proportional to the sine of the phase difference across the SIS
junction. In particular, it is found that the maximum current
$J_{\max}$ decreases when the  mass square $m^2$ of the scalar field
becomes large or the width $L$ of the junction increases. In addition, the condensation
$\langle O\rangle$ dual to the scalar field will also decrease
exponentially with respect to $L$.  The coherence length of the
junction $\xi$ is found very close to each other from the fittings
of the two figures $J_{\max}\sim L$ and $\langle O\rangle\sim L$.
We also analytically study this holographic model by  virtue of the
Sturm-Liouville (S-L) eigenvalue problems~\cite{Siopsis:2010uq}. We
find that the critical chemical potentials $\mu_c$ obtained
analytically are in good agreement with the ones from the numerical
calculation.  Besides, we analytically obtain the critical exponent $1/2$
 from the relation between the operator condensation
 and the chemical potential near the critical point, {\it i.e.},
$\langle O\rangle\propto\sqrt{\mu-\mu_c}$; The charge density $\rho$
is found to be  linearly  proportional to the
chemical potential, {\it i.e.}, $\rho\propto(\mu-\mu_c)$, which is
qualitatively consistent with the numerical calculation.

This paper is organized as follows. In section~\ref{sec2},
we set up  the gravity dual model  of an
inhomogeneous superconductor in a four-dimensional
Einstein-Maxwell-complex scalar theory with a negative cosmological
constant. The holographic insulator/superconductor phase transition
is studied in section~\ref{sec3}. In section~\ref{sec4}, we
numerically solve a set of coupled non-linear equations of motion,
and study the characteristics of the holographic SIS Josephson
junction.  Section~\ref{sec5} is devoted to the conclusions.
In appendix~\ref{Josephsonjunction},  we briefly describe the
Josephson junction, while in appendix \ref{analytical}, we give an
analytical study on the holographic superconductor/insulator phase
transition.

\section{Setup} \label{sec2}

Let us  begin with the action of Einstein gravity with a
cosmological constant in four dimensions:
 \be S=\int d^4x \sqrt{-g}(R-2\Lambda) \;, \ee
where the negative cosmological constant $\Lambda$ is related to
$\ell$ by $\Lambda = -3/\ell^2$, where $\ell$ is the radius of
 AdS space. In this system we have a Ricci flat AdS-Schwarzschild
 black hole solution as
\be\label{matrix1} ds^2 = -\frac{f(r)}{\ell^2} dt^2 + \ell^2
\frac{dr^2}{f(r)} + r^2(dx^2+dy^2) ,
 \ee
where  $f(r)=r^2-{r^3_0}/{r}$ and $r=r_0$ is
the black hole horizon.  With the double Wick rotation, we
can obtain the so-called  AdS soliton solution of this system
\cite{Horowitz:1998ha} \be\label{soliton} ds^2=-r^2 dt^2+\ell^2
\left(r^2-\frac{r^3_0}{r}\right)^{-1}dr^2+r^2dx^2+
\left(r^2-\frac{r^3_0}{r}\right)d\chi^2\;. \ee  Here the
coordinate $\chi$ should have a period with $\beta = 4 \pi \ell/3
r_0$, otherwise there will be a conical singularity at
$r=r_0$. In that case  $r=r_0$ in the metric (\ref{soliton})
becomes a tip of  a cigar-like geometry.   The temperature
associated with this AdS soliton vanishes. In addition, let us
stress here that the coordinate $x$ is infinitely extended, namely
$x\in (-\infty,\infty)$.

 Next we consider the matter sector of the model: a Maxwell field
coupled with a charged complex scalar field. Its action can be
written as
\begin{eqnarray}\label{maxwellaction}
S=\int d^4x\sqrt{-g}\left (
-\frac{1}{4}F^{\mu\nu}F_{\mu\nu}-|\nabla\psi - iA\psi|^2
-m^2|\psi|^2 \right )  ,
\end{eqnarray}
  where $F_{\mu\nu}=\partial_\mu A_\nu-\partial_\nu A_\mu$ is the Maxwell
 field strength and $m$ is the mass of the scalar field.
The equations of motion (EoMs) of  the scalar and Maxwell fields
read  \bea
 \left(\nabla_a - i  A_a \right) \left(\nabla^a - i  A^a\right)\psi - m^2 \psi = 0 \,,\label{ScalarEOM}\\
 \nabla_a F^{ab} - i  \left[\psi^* (\nabla^b - i  A^b)\psi - \psi (\nabla^b + i  A^b)\psi^* \right]= 0 \,.\label{MaxwellEOM}
\eea In order to solve the above equations, we choose an ansatz as
\bea \label{ansz} \psi=|\psi|e^{i\varphi}\;, \qquad A=(A_t, A_r,
A_x,0)\;, \eea where $|\psi|$, $\varphi$, $A_t$, $A_r$, and $A_x$
are all real functions of coordinates $r$ and $x$.
 Instead of $A$, we  work with the gauge-invariant fields: $M=A- d\varphi$.

 We are going to study the holographic model in the probe limit. That
is, we can ignore the back reaction of the matter fields on the AdS
soliton geometry~(\ref{soliton}). In that case, the EoMs for the
matter fields in the AdS soliton background become
\bea 
\label{EO1} \partial_{r}^2|\psi|+\frac{1}{r^2f}\partial_{x}^2|\psi|
    +\left(\frac{f'}{f}+\frac{2}{r}\right)\partial_r|\psi|
    +\frac{1}{f}\left(\frac{M_t^2}{r^2}-fM_r^2
    -\frac{M_x^2}{r^2}-m^2\right)|\psi|&=&0,~~~~~~~\\
\label{EO2} \partial_{r}M_r+\frac{1}{r^2f}\partial_xM_x
 +\frac{2}{|\psi|}\left(M_r\partial_r|\psi|
 +\frac{M_x}{r^2f}\partial_x|\psi|\right)
 +\left(\frac{f'}{f}+\frac{2}{r}\right)M_r&=&0,\\
\label{Mt} \partial_{r}^2M_t+\frac{1}{r^2f}\partial_{x}^2M_t+\frac{f'}{f}\partial_r M_t-\frac{2|\psi|^2}{f}M_t &=&0,\\
\label{Mr} \partial_{x}^2M_r-\partial_{r}\partial_xM_x-2r^2|\psi|^2M_r&=&0,\\
\label{Mx}
\partial_{r}^2M_x-\partial_{r}\partial_xM_r+\frac{f'}{f}\left(\partial_r
M_x-\partial_x M_r\right)-\frac{2|\psi|^2}{f}M_x&=&0,\eea
where the superscript a prime denotes the derivative with respect to $r$. In
the following, we will work with the case $m^2\geq-9/4$, which is
above the Breitenl\"ohner-Freedman
bound~\cite{Breitenlohner:1982bm}. Here we have set
$\ell=1$.

Because Eqs.~\eqref{EO1}-\eqref{Mx} are a set of non-linear coupled
equations, one cannot solve these equations analytically. However,
it is straightforward to solve them numerically. In order to
numerically solve
 Eqs.~\eqref{EO1}-\eqref{Mx}, we need to specify the associated
 boundary conditions in the  $x$ and $r$ directions. By imposing
the Neumann-like boundary condition suggested  in
\cite{Nishioka:2009zj}, we obtain the asymptotic behaviors of
$\psi$, $M_t$, $M_r$ and $M_x$ near the tip $r = r_0$ as \bea
|\psi| &\rightarrow& a_0(x) + a_1(x) (r-r_0)+\mathcal{O}\left((r-r_0)^2\right)\;,\\
M_t&\rightarrow&  b_0(x) + b_1(x) (r-r_0)+\mathcal{O}\left((r-r_0)^2\right)\;,\\
M_r&\rightarrow&  c_0(x) + c_1(x)(r-r_0)+\mathcal{O}\left((r-r_0)^2\right)\;,\\
M_x&\rightarrow&  d_0(x) + d_1(x)(r-r_0)+\mathcal{O}\left((r-r_0)^2\right)\;.
\eea
 Analyzing the EoMs of  the scalar field and Maxwell field on
the AdS boundary at $r=\infty$, we can take the asymptotic forms as
\begin{eqnarray}
|\psi| &\rightarrow&
\frac{\psi^{(1)}(x)}{r^{(3-\sqrt{9+4m^2})/2}}+\frac{\psi^{(2)}(x)}{r^{(3+\sqrt{9+4m^2})/2}}
+\mathcal{O}\left(\frac{1}{r^{(3+\sqrt{9+4m^2})/2+1}}\right)\;,\\
\label{mtexpand} M_t&\rightarrow&\mu(x)-\frac{\rho(x)}{r}+\mathcal{O}\left(\frac{1}{r^2}\right)\;,\\
M_r&\rightarrow&\mathcal{O}\left(\frac{1}{r^3}\right)\;,\\
M_x&\rightarrow&\nu(x)+\frac{J}{r}+\mathcal{O}\left(\frac{1}{r^2}\right)\;
\label{defnu},
\end{eqnarray}
where $\psi^{(i)}$ ($i = 1, 2$) are the condensation values of the
dual scalar operators $\langle\mathcal{O}_{i}\rangle$, according to
the AdS/CFT dictionary. We can set one of $\psi^{(i)}$ to be
vanished. Here we  choose $\psi^{(1)}=0$. In addition, here $\mu$,
$\rho$, $\nu$ and $J$ correspond to  the chemical potential, charge
density, superfluid velocity and current in the boundary field
theory, respectively
\cite{Basu:2008st,Herzog:2008he,Arean:2010xd,Sonner:2010yx,Horowitz:2008bn,Arean:2010zw}.

According to \cite{Nishioka:2009zj}, in the above setup we can
construct a holographic SIS Josephson junction by adjusting
the boundary chemical potential $\mu(x)$: in a region around $x=0$
along the direction $x$, $\mu(x)$ is below the critical chemical
potential $\mu_c$ so that this region behaves like an insulator; while beyond this region $\mu(x)$ is above the
critical chemical potential so that those two regions outside the
insulator region behave like two superconductors. In this
 way we construct a holographic SIS Josephson junction. Thus, the gauge invariant phase difference $\gamma$ across
the insulator region can be defined as~\cite{Horowitz:2011dz} \be
\gamma=- \int_{-\infty}^\infty dx \;[\nu(x) -\nu(\pm\infty)]\;.
\label{gamma} \ee
 The Josephson current $J$ across the Josephson junction has a sine relation
to the phase difference $\gamma$. We will numerically confirm this
in Sec.~\ref{sec4}.

\section{Holographic superconductor/insulator phase transition}\label{sec3}

In this section, we will first analyze the critical chemical
potential $\mu_c$ for the superconductor/insulator phase transition
near $x\rightarrow\infty$, in order to give a proper configuration
of the chemical potential for the SIS Josephson junction in the next
section. In the limit $x\to\infty$, we assume that all the fields
are independent of $x$, {\it i.e.}, all the fields are homogeneous
in the $x$-direction. Therefore, Eqs. \eqref{EO1}-\eqref{Mx}
can be reduced to two coupled ordinary differential equations for
$|\psi|$ and $\phi$ as
 \bea
\partial_{r}^2|\psi|+\left(\frac{f'}{f}+\frac{2}{r}\right)\partial_r|\psi|+\frac{1}{f}\left(\frac{M_t^2}{r^2}-m^2\right)|\psi|&=0\;,\label{EO1o}\\
\partial_{r}^2M_t+\frac{f'}{f}\partial_r M_t-\frac{2|\psi|^2}{f}M_t &=0\;.\label{Mto}
\eea
 Here we will numerically solve the
 EoMs \eqref{EO1o} and \eqref{Mto}, while some analytical results will be presented in
 appendix \ref{analytical}, for  comparison.
  We will see that the critical chemical potentials obtained
 from the two approaches are very close to each other.

 The asymptotic forms
of $|\psi|$ and $M_t$ near the boundary $r=\infty$ are \bea
|\psi| &= &\frac{\psi^{(1)}}{r^{(3-\sqrt{9+4m^2})/2}}+\frac{\psi^{(2)}}{r^{(3+\sqrt{9+4m^2})/2}} ,\\
    M_t &= &\mu-\frac{\rho}{r} ,
\eea
 in which the four  parameters ($\psi^{(1)}, \psi^{(2)}, \mu, \rho$) are constants independent
of $x$, $\mu$ and $\rho$ are the chemical potential and charge
density, respectively, while $\psi^{(1)}$ and $\psi^{(2)}$ are the
condensations of dual operators in the superconducting phase.

\begin{figure}[]
\begin{center}
 \subfigure[]{
  \includegraphics[scale=0.58]{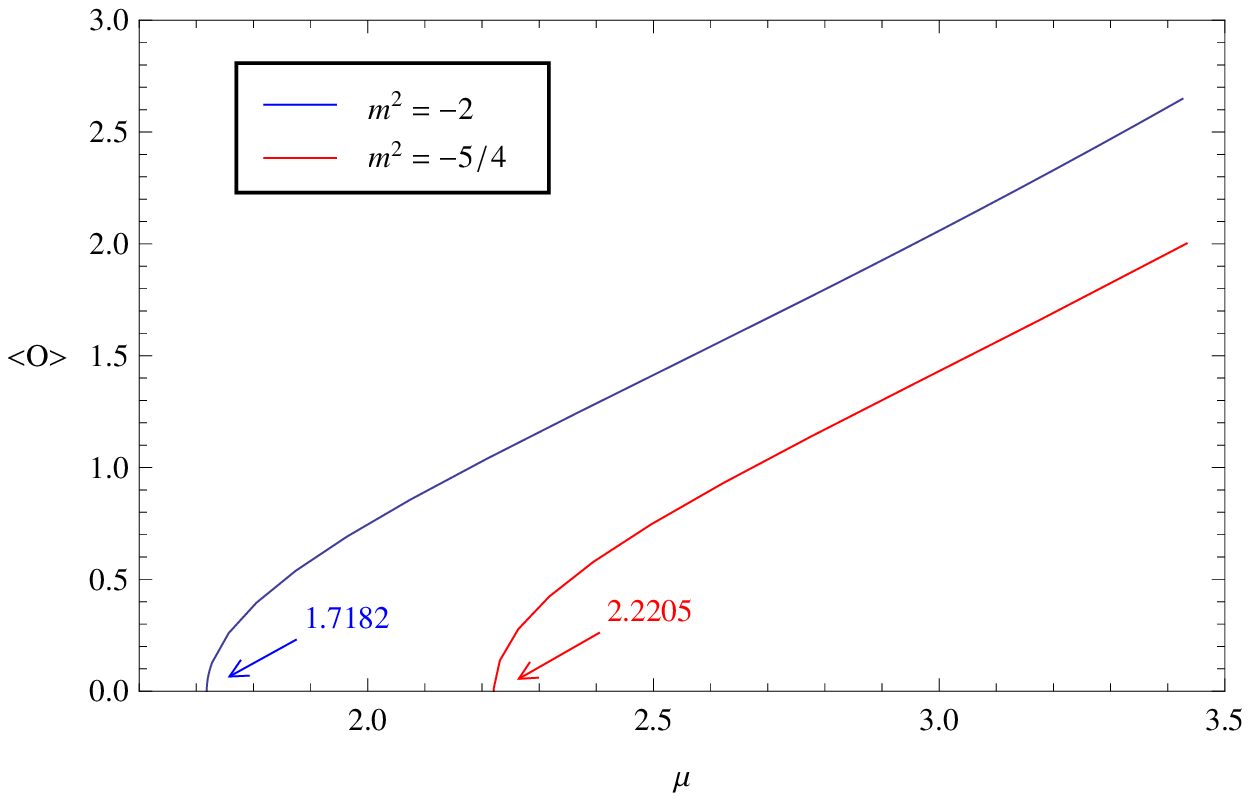}}\label{fig11}
 \subfigure[]{
  \includegraphics[scale=0.57]{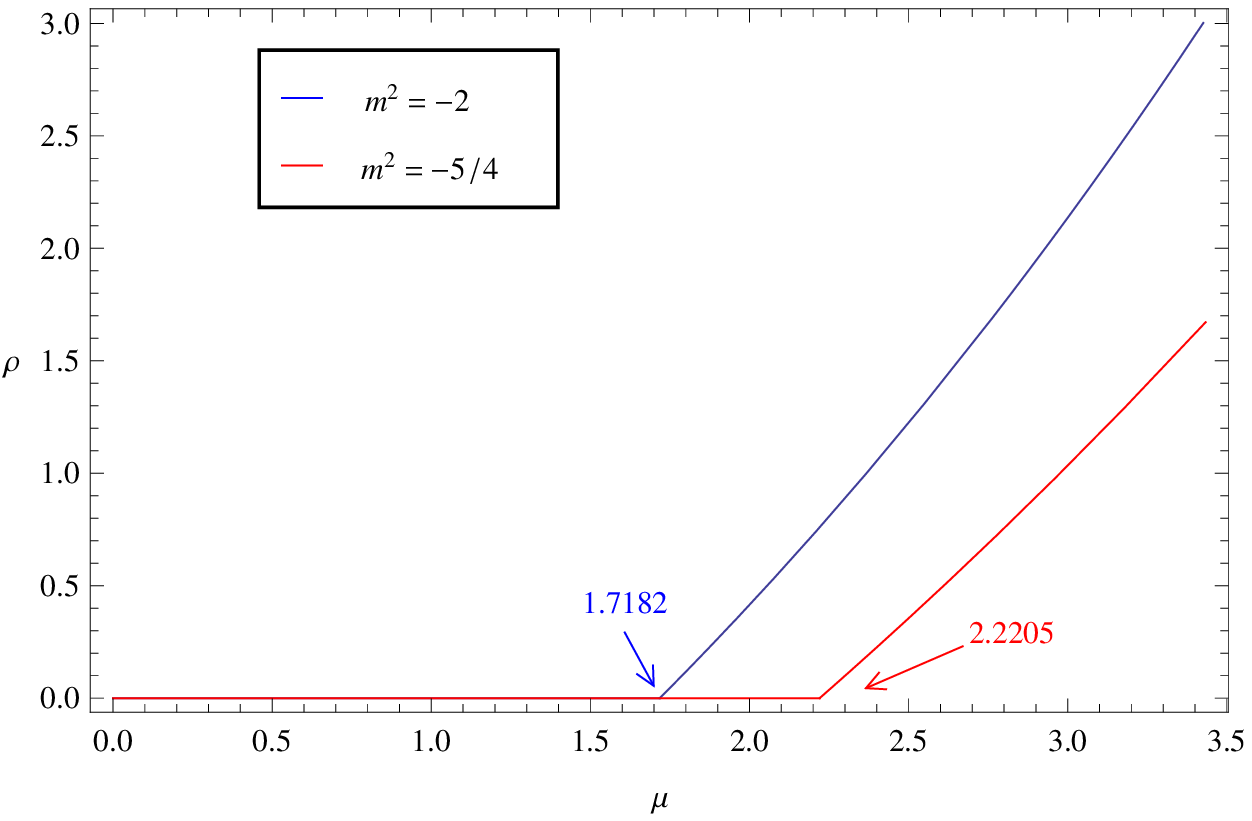}}\label{fig12}
\end{center}
\caption{(a) The condensation of the operator $\langle O\rangle $
versus the chemical potential $\mu$; (b) The charge density $\rho$
versus the chemical potential $\mu$. }
 \label{fig_scalar_warpfactor}\label{fig1}
\end{figure}

 In the numerical calculations, it is convenient to use the
shooting method to solve the boundary value problem for the ordinary
differential equations (\ref{EO1o}) and (\ref{Mto}). In the
following  we will set $r_0=1$ and $\ell=1$. In Fig.~\ref{fig1} we
plot the condensation of the operator
 $\langle \mathcal{O}\rangle$ and charge density $\rho$  as functions of the chemical potential $\mu$.
 We can see that the charged scalar operator condensates when the chemical potential is
  above  critical value $\mu_c$:
\bea \label{muc} \mu_c &\approx& 1.7182\quad \text{for} \quad m^2=-2, \\
    \label{muc54} \mu_c &\approx& 2.2205\quad \text{for} \quad m^2=-5/4.\eea
We can see that the critical chemical potential
$\mu_c$ grows when $m^2$ increases, which means that the
condensation becomes hard when $m^2$ increases.  In the case of $\mu
< \mu_c$,  the condensation does not happen and the charge
density $\rho$ is always vanishing,
  one realizes a holographic insulator phase; while $\mu
> \mu_c$, the charged operator condensates, which  breaks the U(1)
gauge symmetry spontaneously. Therefore, when the chemical
potential crosses the critical value $\mu_c$,  one can realize a
holographic insulator/superconductor phase transition. In appendix
\ref{analytical}, we  get analytically the values of the critical
chemical potential by using the S-L eigenvalue procedure, which are
very close to the ones from the numerical calculations.

In addition, near the critical point, we can read off the critical
behavior of the condensation $\langle \mathcal{O}\rangle$ and the
charge density $\rho$ from Fig.~\ref{fig1} as \bea
  \label{Omuc} \langle \mathcal{O}\rangle &\approx&
  1.3134\sqrt{\mu-\mu_c},\quad
  \rho \approx  1.4229(\mu-\mu_c) \quad \text{for}\quad m^2=-2, \\
   \label{Omuc54}\langle \mathcal{O}\rangle &\approx&
   1.4231\sqrt{\mu-\mu_c},\quad
   \rho \approx 1.2758(\mu-\mu_c)\quad \text{for}\quad m^2=-5/4.
\eea
 We can see that $\langle \mathcal{O}\rangle$ is proportional to the
 square root of $\mu-\mu_c$, and $\rho$ is linearly  proportional to
 $\mu-\mu_c$. We also obtain these square root and linear relations in the
 analytical study of the critical behavior in appendix
 \ref{analytical}.

\section{Numerical solutions of the holographic SIS Josephson junction}\label{sec4}

In this section, we will numerically study  the holographic SIS Josephson
junction. For this aim, following \cite{Horowitz:2011dz}, we
 choose the profile of the chemical potential $\mu(x)$ as
\be\label{profile}
\mu(x)=\mu_\infty\left\{1-\frac{1-\epsilon}{2\tanh(\frac{L}{2\sigma})}
  \left[\tanh\left(\frac{x+\tfrac{L}{2}}{\sigma}\right)
  -\tanh\left(\frac{x-\tfrac{L}{2}}{\sigma}\right)\right]\right\} ,
\ee where $\mu_{\infty}\equiv\mu(\infty)=\mu(-\infty)$ is the value
of the chemical potential $\mu$ at $x=\pm\infty$, and
  the parameters $L$, $\sigma$ and $\epsilon$ are the width,  steepness and depth of the junction, respectively.
  Actually, we can suitably choose the parameters in $\mu(x)$  so
  that
  \bea \left\{\begin{array}{ll}
               \mu(x)<\mu_c, & \text{for}~ -L/2<x<L/2, \\\\
               \mu(x)>\mu_c, & \text{for}~ x<-L/2 ~\text{and}~
               x>L/2.
             \end{array}\right.
             \eea
Thus the central
  part of the junction is a holographic insulator while the two sides are two holographic superconductors.
   Such a junction is an SIS Josephson
  junction as we expected.

With the above chemical potential, we can numerically solve the set of coupled equations \eqref{EO1}-\eqref{Mx}
    by virtue of the spectral method.
\begin{figure}[]
  \begin{center}
  \includegraphics[scale=0.7]{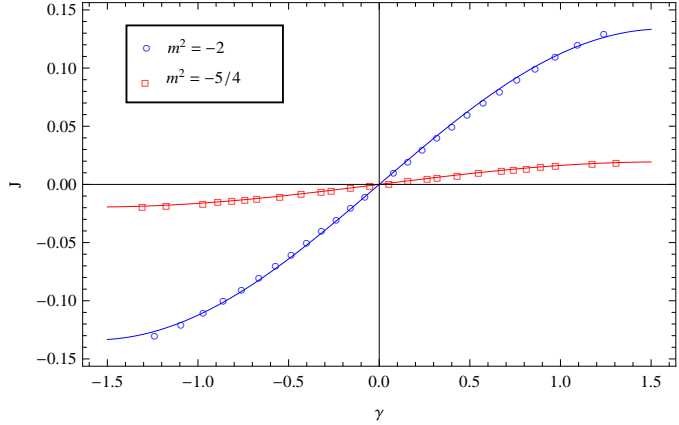}
  \end{center}
  \caption{The sine relation between the Josephson current $J$
  and the phase difference $\gamma$ for two different $m^2$. Here the parameters are set to
  $\mu_\infty=4$, $L=3$, $\epsilon=0.4$, $\sigma=0.5$.
  }\label{sine}
\end{figure}
In Fig.~\ref{sine}, we plot the current $J$  as a function of the
phase difference $\gamma$ with $\mu_\infty=4$, $L=3$,
$\epsilon=0.4$, $\sigma=0.5$. The circles and the squares are from
the numerical calculations while the solid lines are the fittings of
them.  The relation between $J$ and $\gamma$ can be fitted as
 \bea
 J&\approx& 0.1336 \sin \gamma,\quad \text{for} \quad m^2=-2, \\
 J&\approx& 0.0194\sin
\gamma,\quad \text{for}\quad m^2=-5/4. \eea
We see that the Josephson
current $J$ is indeed proportional to the sine of the phase
difference $\gamma$, which realizes the SIS Josephson junction.
Further we  find that the amplitude of the sine relation or the
 maximum current $J_{\max}$ decrease when $m^2$ becomes
 large.
\begin{figure}\label{fig2}
\begin{center}
 \subfigure[]{\label{figphi}
  \includegraphics[width=0.45\textwidth]{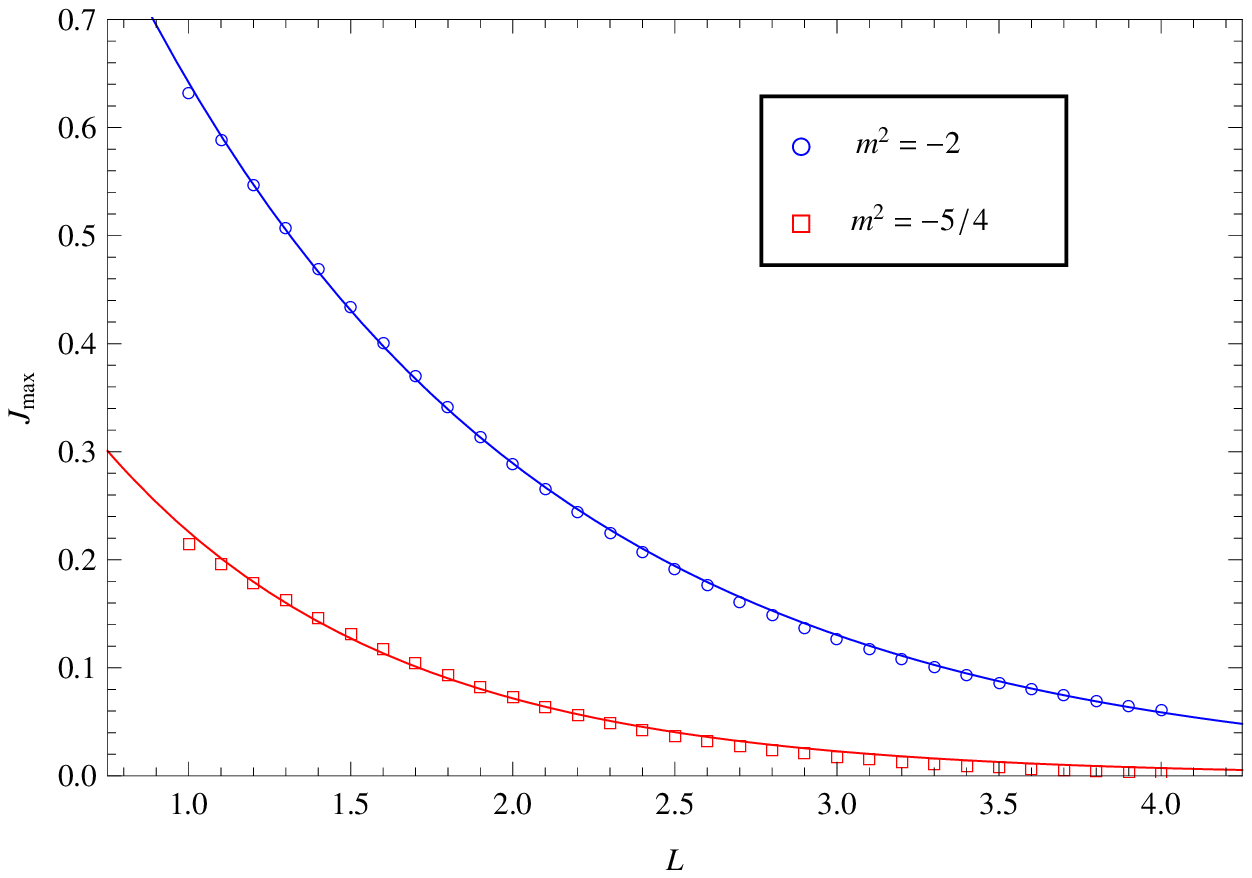}}
\hspace{0.5cm}
 \subfigure[]{\label{figphiprime}
  \includegraphics[width=0.47\textwidth]{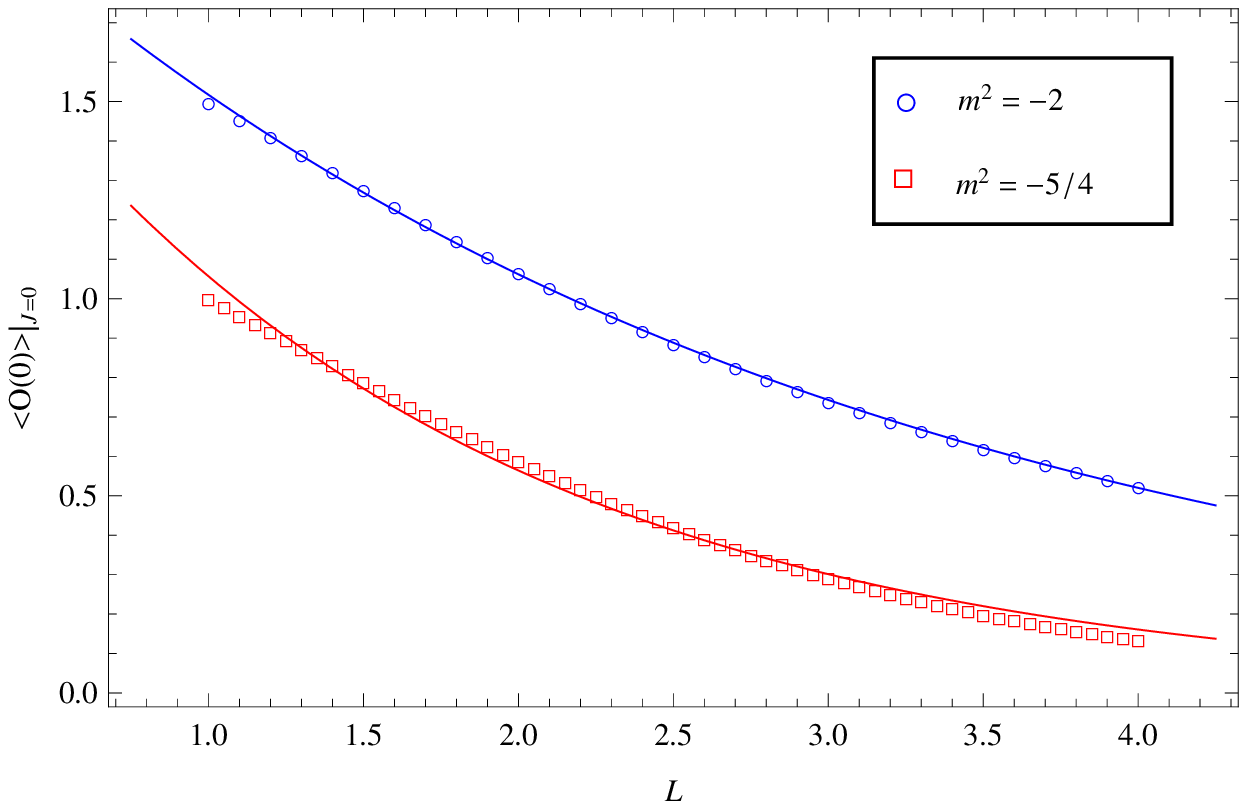}}
\end{center}  \vskip -5mm
\caption{ (a) The maximal current $J_{\max}$ versus the width $L$ of
the junction; (b) The condensation $\langle
\mathcal{O}(0)\rangle|_{J=0} $ versus the width $L$ of the junction,
for two different $m^2$.
 Here the parameters are chosen as $\mu_\infty=4$, $\epsilon=0.4$, and $\sigma=0.5$.}
 \label{fig_scalar_warpfactor}
\end{figure}
The relation between the maximum current $J_{\max}$ and the width of
the junction $L$ is plotted in Fig.~\ref{figphi}; And the relation
between the condensation $\langle\mathcal{O}(0)\rangle=\psi^{(2)}$
at zero current and $L$ is given in Fig. \ref{figphiprime}. The
circles and squares are obtained from the numerical calculations
while the solid lines are the fittings of them. From appendix
\ref{Josephsonjunction}, one learns that the dependencies of $J_{\max}$
and $\langle \mathcal{O}(0)\rangle|_{J=0}$ on $L$ behave like
\begin{eqnarray}
J_{\max} &=& A_0\,e^{-\frac{L}{\xi}},\label{jmaxell} \\
\langle \mathcal{O}(0)\rangle|_{J=0}&=&
A_1\,e^{-\frac{L}{2\,\xi}},\label{oell}
\end{eqnarray}
where $\xi$ is the coherence length of the SIS Josephson junction,
while $A_0$ and $A_1$ are two constants. In our holographic model,
the fitted result in Fig.~\ref{figphi} gives
  {
 \bea
 J_{\max}&\approx& 1.4230 e^{-{L}/{1.2552}},\quad
 \text{for}\quad m^2=-2,\\
  J_{\max}&\approx& 0.7107
 e^{-{L}/{ 0.8720}},\quad
 \text{for} \quad m^2=-5/4,
 \eea
 }
 while in Fig. \ref{figphiprime} the fitted result has the forms
  {
 \bea
 \langle O\rangle&\approx&2.1671 e^{-L/(2\times 1.4015)},\quad \text{for}\quad
 m^2=-2,\\
 \langle O\rangle&\approx&1.9783 e^{-L/(2\times 0.7969)},\quad \text{for}\quad
 m^2=-5/4.
 \eea
 }
The coherence lengths $\xi$ fitted from the two plots are consistent
to each other within $10.5\%$ error. Thus we see that in the
holographic model, the familiar results (\ref{jmaxell}) and
(\ref{oell}) for the SIS Josephson junction can be indeed reproduced on the
gravity side.

\begin{figure}
  \begin{center}
  \includegraphics[scale=0.7]{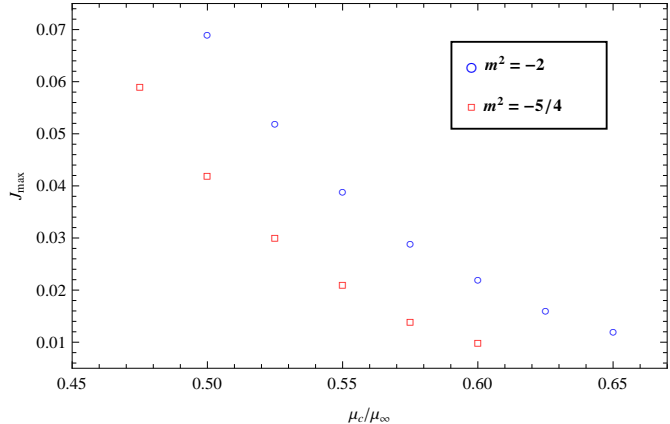}
  \end{center}
  \caption{The  maximum
   current $J_{\max}$ versus $\mu_c/\mu_\infty$ for two
   different $m^2$. The parameters are set to $L=3$, $\epsilon=0.4$, $\sigma=0.5$.}\label{fig4}
\end{figure}

We also plot the relation between the maximum current $J_{\max}$ and
$\mu_c/\mu_\infty$ in Fig.~\ref{fig4}, from which one can see that
$J_{\max}$ will decrease when $\mu_c/\mu_\infty$ increases, and as
$\mu_c/\mu_\infty$ goes close to $1$, the current  is suppressed
very much.  This result is expected, because when
$\mu_c/\mu_{\infty} \to 1$, the chemical potential in both sides of
the insulator just reaches the critical value. In this case the
superconducting current is very small and  the ``strength" of their
superconductivity is rather weak. We expect that the current
vanishes finally at $\mu_c/\mu_\infty = 1$. On the other hand, we
can see the maximal current quickly increases when
$\mu_c/\mu_\infty$ becomes small. This tendency is also understandable,
since in this case the two sides of the junction are in the
superconducting phases far from the critical point, and the
``strength" of the superconductivity becomes very strong.

\section{Conclusions}\label{sec5}
In this paper, we studied the holographic
insulator/superconductor phase transition in a four-dimensional AdS
soliton background. It was found that the critical chemical
potential $\mu_c$ increases when the mass square $m^2$ of the scalar
field grows.  We  investigated the behaviors of the
condensation $\langle \mathcal{O}\rangle$ and the charge density
$\rho$, and found that  the numerical results near the critical point
are in good agreement with those  obtained from an analytical
 procedure.

 Furthermore,  with the help of the holographic
superconductor/insulator model, suitably  setting the profile of the
chemical potential $\mu(x)$ on the AdS boundary, and following
\cite{Horowitz:2011dz}, we constructed a holographic model for an
SIS Josephson junction. We  numerically solved a set
 of non-linear equations of motion and confirmed
that the Josephson current $J$ of the SIS junction
 is indeed  proportional to the sine of the phase difference $\gamma$ across the
junction. Besides,  we  also shown that the maximum
current $J_{\max}$
 decreases when the mass square of the scalar field  becomes large.  We  further studied
the dependence of $J_{\max}$ and $\langle
\mathcal{O}(0)\rangle|_{J=0}$ on the width $L$ of the junction, and
found that $J_{\max}$ and $\langle \mathcal{O}(0)\rangle|_{J=0}$
  are suppressed exponentially when $L$ increases.
Thus we reproduced these main features of Josephson junction on the
gravity side. In addition, $J_{\max}$ was found to be a decreasing
function of $\mu_c/\mu$.

 The Josephson effect is an important phenomenon associated
with superconductor. Some features  of Josephson junction are robust
such as the sine relation between the current and the phase
difference across the junction. For the holographic models of the
SNS Josephson junction constructed in \cite{Horowitz:2011dz} and
 the SIS Josephson junction discussed in this paper, one
can indeed confirm this relation on the gravity side.
 Therefore, it is interesting to further study the
holographic model for the
superconductor-superconductor-superconductor junction,  where the
middle superconductor is different from the other two
superconductors. We expect that the sine relation  will also hold in
that case. Finally let us mention here that it is also of great
interest to investigate the effect of magnetic field on the
Josephson junction from the holographic point of
view~\cite{Rowell1963}.

\acknowledgments

S.T and H.Q.Z would like to thank the hospitalities of Lanzhou
University. Y.Q.W and Y.X.L were supported in part by the National
Natural Science Foundation of China (No. 11005054 and No. 11075065),
and in part by the Fundamental Research Funds for the Central
Universities (No. lzujbky-2012-18 and No. lzujbky-2012-k30 ).  R.G.C
and H.Q.Z were supported in part by the National Natural Science
Foundation of China (No.10821504, No.10975168 and No.11035008), and
in part by the Ministry of Science and Technology of China under
Grant No. 2010CB833004.

\appendix

\section{Brief description of Josephson junction}
\label{Josephsonjunction}
 To be complete, we will briefly review the
Josephson junction in this {appendix}. Considering two superconductors
separated by an insulating barrier, in 1962,
Josephson~\cite{Josephson:1962zz} made a remarkable prediction that
there exists a current flowing across the middle insulator even when
there is no voltage difference between these two superconductors.
This supercurrent is caused by quantum tunneling of Cooper pair.
This is just the Josephson effect. The current is related to the
difference of phases of macroscopic wave functions in the two
superconductors:
\begin{equation}
\label{a1}
 J= J_{\max} \sin \gamma,
\end{equation}
where $J_{\max}$ is the maximal current (critical current) and
$\gamma$ is the phase difference of the macroscopic wave function
$\psi=|\psi|e^{i \theta}$ in two superconductors, which is
introduced as an order parameter in the Ginzburg-Landau model. This
remarkable prediction was experimentally confirmed
soon~\cite{Anderson}.

The Josephson's prediction was based on a microscopic theoretical
analysis of quantum mechanical tunneling of electrons through the
insulating barrier layer. But in fact, this Josephson effect is much
more general, it occurs when two superconductors are connected by a
weak link, the latter is called Josephson junction.  The weak link
can be an insulator as Josephson originally proposed, or a normal
metal layer, or simply a short and narrow constriction. In those cases, the Josephson junction are called SIS,
SNS and SCS junctions, respectively.

Suppose that $\psi_1=|\psi_1|e^{i\theta_1}$ and
$\psi_2=|\psi_2|e^{i\theta_2}$ are two macroscopic wave functions
describing these two superconductors. Due to the weak coupling
between them, the two wave functions are not independent of each
other. Instead they relate to each other as:
{ $\partial_x \psi_1 =   K \psi_2$  and $ \partial_x \psi_2 = - K \psi_1$},
where $K$ is a parameter representing the strength of the weak
coupling, and $x$ is the direction perpendicular to the boundary
surface between superconductor and insulator. In the case with two
 { identical} superconductors, one  has
$|\psi_1|=|\psi_2|\equiv |\psi|=\sqrt{\rho_0}$, where $\rho_0$ is
the density of Cooper pair.
 Generally speaking,  the supercurrent associated with a macroscopic
wave function $\Psi(\mathbf{r})$ is given as $\mathbf{J} = -i
\frac{e^*\hbar}{2m^*} \left( \Psi^*  \nabla \Psi - \Psi  \nabla
\Psi^* \right) $, where $e^*$ and $m^*$ mean the charge and the mass
of the Cooper pair.
From this, one can obtain the Josephson current flowing over the
weak link by substituting $\psi_1=|\psi|e^{i \theta_1}$ and
$\psi_2=|\psi|e^{i \theta_2}$ into the supercurrent equation as
\begin{eqnarray}
J_x = J_{\max} \sin (\theta_2 -\theta_1)=J_{\max} \sin \gamma,
\end{eqnarray}
along the direction $x$, where  $\displaystyle J_{\max} \equiv K
\frac{e^* \hbar}{m^*}\rho_0 $ is the maximal  Josephson current.

The maximal Josephson current $J_{\rm max}$ is closely related to
the width $L$ of the weak link (see, for example, \cite{tinkham}) as
\begin{equation}
\label{a4}
 J_{ \max} \sim \rho_0 \,e^{-L/\xi},
\end{equation}
where the coherence length $\xi$, which  is  one of the characteristic scales
 of superconductors, can vary with superconducting materials.  The
maximal current is also proportional to the density of Cooper pair
in the weak link as $J_{\max} \sim |\psi|^2$.  Considering the
relation between the number density $\rho$ and the condensation
$\langle \mathcal{O} \rangle$: $\rho \sim \langle \mathcal{O}
\rangle^2$, see, for example, (\ref{rhomuc2}), one has
\begin{equation}
\label{a5}
 \langle \mathcal{O} \rangle \sim J_{\max}^{1/2} \sim
e^{-L/2\xi},
\end{equation}
in the weak link of the Josephson junction. Indeed, in our
holographic model of SIS Josephson junction, we have reproduced the three
relations (\ref{a1}), (\ref{a4}) and (\ref{a5}).

\section{Analytical study on the superconductor/insulator phase transition}
\label{analytical}

In this appendix we will take advantage of the S-L method
\cite{Siopsis:2010uq} to analytically study the homogenous solutions
 of the equations of motion near the critical point of the phase
transition. For further applications of the S-L method to other
holographic superconductor models, see Refs.
\cite{Zeng:2010zn,Li:2011xja,Cai:2011ky,Pan:2011ah,Momeni:2011iw,Cai:2011tm,Lee:2012qn}.

\subsection{Analytical results of the critical chemical potential for
general mass}
\begin{figure}[]
  \begin{center}
  \includegraphics[scale=0.7]{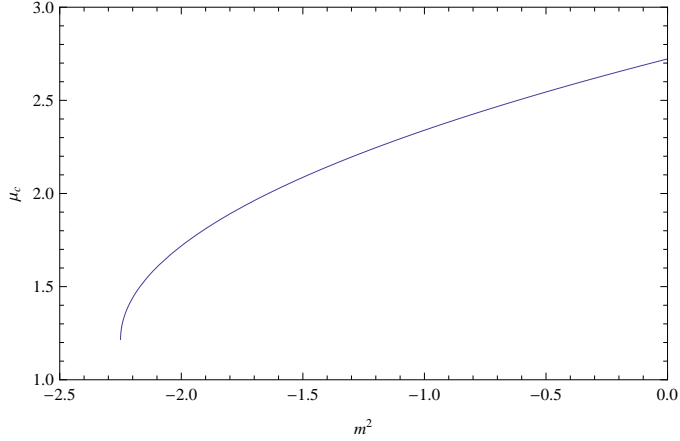}
  \end{center}
\caption{The critical chemical potential $\mu_c$ as a function of
the mass square $m^2$. { The plot is obtained from minimizing the
formula \eqref{mu2} for certain $\alpha$.}} \label{mum}
\end{figure}

In this case all matter fields are independent of the coordinate
$x$. By setting $z=1/r$, the EoMs \eqref{EO1} and \eqref{Mt}  are
reduced to (we have set $M_t=\phi$ here)
 \bea
 \label{psiz}
 \partial_z^2\psi(z)-\frac{2+z^3}{z-z^4}\partial_z\psi(z)+\frac{\phi(z)^2z^2-m^2}{z^2(1-z^3)}\psi(z)&=&0, \\
 \label{phiz}
 \partial_z^2\phi(z)-\frac{3z^2}{1-z^3}\partial_z\phi(z)-\frac{2\psi(z)^2}{z^2(1-z^3)}\phi(z)&=&0.
 \eea
At the critical point, the scalar field $\psi$ vanishes. Near the
critical point, we can introduce a trial function $F(z)$ into the
asymptotical form of $\psi$ near the boundary like
 \be\label{ffunction} \psi|_{z\rightarrow0}\sim\langle O\rangle z^\Delta F(z),
 \ee
 where $\Delta \equiv 3/2+\sqrt{9/4+m^2}$ is the conformal
dimension of the dual operator. It is easy to see from the boundary
conditions of $\psi$ that $F(0)=1$. For simplicity, we can set
$F~'(0)=0$. Therefore, the simplest form of $F(z)$ is $F(z)=1-\alpha
z^2$, where $\alpha$ is a constant. Besides, near the critical point
of the phase transition, $\phi(z)\sim \mu$. Thus, substituting
formula \eqref{ffunction} into Eq. \eqref{psiz}, and multiplying
$K(z) \equiv z^{-2(1- \Delta)} (-1 + z^3)$ to both sides, we can
reach
  \begin{eqnarray}
\partial_z \bigg( K(z)\partial_z F(z)\bigg)
+ \bigg( -P(z) + Q(z) \mu^2 \bigg)F(z) =0,
\end{eqnarray}
 with
\begin{eqnarray}
P(z) &\equiv& -z^{-2(2 - \Delta)} \left[ m^2 +\Delta \left\{
3-\Delta(1 - z^3)\right\} \right] ~{\rm and}~ Q(z) \equiv
-z^{-2(1-\Delta)}.
\end{eqnarray}
 Through the S-L eigenvalue method, we can obtain the
critical chemical potential by minimizing the following functional
at some  value of $\alpha$,
 \be
 \label{mu2}
 \mu^2=
 \frac{\displaystyle \int_0^1 dz \left\{K(z) (\partial_z F(z))^2 + P(z) F(z)^2\right\}}{\displaystyle \int_0^1 dz ~
 Q(z)F^2(z)}.\ee
 Fig.\ref{mum} plots the relation
 between $\mu_c$ and $m^2$. In particular, the critical chemical potentials for the cases of $m^2=-2$ and $-5/4$
 are, respectively,
\begin{eqnarray}\label{mucsl2}
&&\mu_c \approx 1.71884  \quad \textrm{for} \quad m^2= -2,\\
\label{mucsl54}&&\mu_c \approx 2.22116  \quad \textrm{for} \quad
m^2= -5/4.
\end{eqnarray}
We see that these
 values \eqref{mucsl2} and \eqref{mucsl54}  of the critical chemical potential are in good
 agreement with the numerical results \eqref{muc} and \eqref{muc54},
 respectively.

 \subsection{Analytical relations of $\langle O\rangle\sim(\mu-\mu_c)$ and $\rho\sim(\mu-\mu_c)$}

 Near the critical point $\mu\rightarrow\mu_c$, the condensation
value of the operator $\langle O\rangle$ is very small, therefore,
we can expand $\phi(z)$ in the series of $\langle O\rangle$ like
 \be\label{phiexpand} \phi\sim\mu_c+\langle O\rangle\zeta(z),\ee
where, $\zeta(z)$ is a coefficient function independent of $\langle
O\rangle$. We can easily find that the behavior of $\phi(z)$ near
the tip of the background geometry imposes \be\label{xihorizon}
\zeta(1)=0.\ee
 The EoMs of $\zeta$ can be  obtained from  Eqs. \eqref{phiz}, \eqref{ffunction}
 and
 \eqref{phiexpand} by taking the leading order of
 $\langle O\rangle$ as
 \be \label{xieom}
 \partial_z^2\zeta(z)-\frac{3z^2}{1-z^3}\partial_z\zeta(z)=\frac{2\mu_c\langle
 O\rangle z^{2\Delta-2}F(z)^2}{1-z^3}.\ee
Multiplying $T(z)=-1+z^3$ to both sides of \eqref{xieom}, we arrive
at
 \be \frac{d}{dz}\bigg(T(z)\partial_z\zeta(z)\bigg)=-2\mu_c\langle
 O\rangle z^{2\Delta-2}F(z)^2.\ee
 Integrating both sides of the above
equation, we can get the Neumann boundary condition of $\zeta$ near
$z\to 0$ as
 \bea \label{xibrd}
 (-1+z^3)\partial_z\zeta(z)\big|_0^1
  &=&\partial_z\zeta(z)\big|_{z\rightarrow 0}\nonumber \\
  &=&-2\mu_c\langle O\rangle\int_0^1dz ~z^{2\Delta-2}F(z)^2 \nonumber \\ &=&-2\mu_c\langle
 O\rangle\bigg(\frac{1}{2\Delta-1}-\frac{2\alpha}{2\Delta+1}+\frac{\alpha^2}{2\Delta+3}\bigg).\eea
Therefore, from  Eq.~\eqref{xieom} and the boundary conditions
\eqref{xihorizon} and \eqref{xibrd} of $\zeta$, we can obtain
 \bea \label{xi}
 \zeta(z)&=&\frac{1}{9} \mu_c\langle O\rangle \bigg\{-\frac{18
\alpha ^2 z^{2 \Delta +1}}{4 \Delta ^2+8 \Delta
   +3}-\frac{9 z^{2 \Delta }}{\Delta -2 \Delta ^2}\nonumber\\&&+ \frac{18
\alpha ^2 z^{2 \Delta +1} }{4 \Delta ^2+8 \Delta +3}\,
   _2F_1\left(1,\frac{1}{3} (2 \Delta +1);\frac{2 (\Delta
   +2)}{3};z^3\right)\nonumber\\&&-\frac{18
   \alpha  z^{2 \Delta +2}}{2 \Delta
   ^2+3 \Delta +1} \, _2F_1\left(1,\frac{2 (\Delta
   +1)}{3};\frac{1}{3} (2 \Delta +5);z^3\right)\nonumber\\&&+\frac{18 z^{2 \Delta +3} }{4 \Delta ^2+4 \Delta -3}
   \,
   _2F_1\left(1,\frac{2 \Delta }{3}+1;\frac{2 \Delta
   }{3}+2;z^3\right)\nonumber\\&&+ \frac{1}{\Delta  (2 \Delta -1) (2 \Delta +1) (2
   \Delta +3)}\bigg[6 \alpha ^2 \Delta  \left(4 \Delta ^2-1\right) \psi
   ^{(0)}\left(\frac{2 \Delta }{3}+\frac{1}{3}\right)\nonumber\\&&+2
   \alpha ^2 \Delta  (2 \Delta -1) \left(\sqrt{3} \pi  (2
   \Delta +1)+9\right)-4 \sqrt{3} \pi  \alpha  \Delta
   \left(4 \Delta ^2+4 \Delta -3\right)\nonumber\\&&-12 \alpha  \Delta
   \left(4 \Delta ^2+4 \Delta -3\right) \psi
   ^{(0)}\left(\frac{2 (\Delta +1)}{3}\right)+\left(2
   \sqrt{3} \pi  \Delta -9\right) \left(4 \Delta ^2+8
   \Delta +3\right)\nonumber\\&&+6 \Delta  \left(4 \Delta ^2+8 \Delta
   +3\right) \psi ^{(0)}\left(\frac{2 \Delta
   }{3}+1\right)+3 \Delta  \bigg(\alpha ^2 \left(4 \Delta
   ^2-1\right)+\alpha  \left(-8 \Delta ^2-8 \Delta
   +6\right)\nonumber\\&&+4 \Delta ^2+8 \Delta +3\bigg) \left(-\log
   \left(z^2+z+1\right)+2 \log (1-z)-2 \sqrt{3} \tan
   ^{-1}\left(\frac{2 z+1}{\sqrt{3}}\right)\right.\nonumber\\&&+2 \gamma +\log
   (27)\bigg)\bigg]
\bigg\},
   \eea
where, $\,_2F_1(a,b;c,d)$ is the hypergeometric function,
$\psi^{(0)}$ is the digamma function and $\gamma$ is the Euler gamma
function.

 Near the boundary $z=0$, we can expand $\phi(z)$ from \eqref{mtexpand} and \eqref{phiexpand}
 as
  \be\label{phibrd} \phi(z)\sim \mu-\rho z\sim \mu_c+\langle
  O\rangle\bigg(\zeta(0)+\zeta'(0)z+\frac12\zeta''(0)z^2+\cdots\bigg).\ee
 Therefore, comparing the coefficients of the $z^0$ term of \eqref{phibrd} we
  have
  \bea \mu-\mu_c\sim\langle O\rangle\zeta(0)\propto\langle O\rangle^2.\eea
   From the critical values of $\mu_c$ and the corresponding $\alpha$ above, we can
   reach
 \bea\label{Omuc2} \langle O\rangle &\approx& 1.08790\sqrt{\mu-\mu_c},\quad \text{for}\quad m^2=-2,\\
                  \langle O\rangle &\approx& 1.09588\sqrt{\mu-\mu_c},\quad \text{for}\quad
                  m^2=-5/4.\eea
 which are qualitatively consistent with the numerical results in \eqref{Omuc} and \eqref{Omuc54}. The
 critical exponent $1/2$ between the condensation value and the
 chemical potential is also consistent with the one from  the mean field theory.
Besides, we notice from the $z^1$ terms of both sides of
\eqref{phibrd} that
 \be\label{rhomuc2} \rho\sim-\langle O\rangle\zeta'(0)\propto\langle
 O\rangle^2\propto(\mu-\mu_c).\ee
This linear relation between the charge density and the chemical
potential is also qualitatively consistent with the numerical
results in \eqref{Omuc} and \eqref{Omuc54}. From \eqref{rhomuc2} we
obtain
\bea \rho&\approx& 1.05323(\mu-\mu_c),\quad \text{for}\quad m^2=-2,\\
      \rho&\approx& 0.95214(\mu-\mu_c),\quad \text{for}\quad
      m^2=-5/4.\eea
 Here the discrepancy in the numerical factors from both methods in
 the relations $\langle O\rangle \sim \sqrt{\mu-\mu_c}$ and  $\rho\sim (\mu-\mu_c)$
might be due to the fact that we have only adopted a simplest form
of the trial function $F(z)$ in \eqref{ffunction}, if we introduce
higher order terms of $z$ into the trial function, this may improve
the  discrepancy of the numerical factors.

\end{document}